# Effects of asymmetric splayed columnar defects on the anomalous peak effect in $Ba_{0.6}K_{0.4}Fe_2As_2$

Ayumu Takahashi[1], Sunseng Pyon[1], Tadashi Kambara[2], Atsushi Yoshida[2], and Tsuyoshi Tamegai[1]

1 Department of Applied Physics, The University of Tokyo, 7-3-1 Hongo, Bunkyo-ku, Tokyo 113-8656, Japan
2 Nishina Center, RIKEN, 2-1 Hirosawa, Wako, Saitama 351-0198, Japan

**Abstract**

In order to enhance the critical current density ($J_c$) of superconductors, introduction of columnar defects (CDs) through swift-particle irradiation is effective. By dispersing the direction of CDs (splayed CDs), not only further enhancement of $J_c$ has been confirmed but also an anomalous peak effect (APE) in $J_c$ at a certain magnetic field determined by the irradiation dose was observed. It has been proposed that the APE arises from the suppression of kink motion and/or the effects of vortex entanglement in systems with splayed CDs. In this study, we measure $J_c$ properties of optimally K-doped Ba-122-type iron-based superconductor $Ba_{0.6}K_{0.4}Fe_2As_2$ single crystals with splayed CDs that are introduced asymmetrically with respect to the $c$-axis by irradiating 2.6 GeV U ions. We discuss the significance of the average direction of splayed CDs to the APE. We also discuss the relationship between the APE and the entanglement of vortices.

# 1. Introduction

Introduction of columnar defects (CDs) to superconductors through swift-particle irradiation enhances their critical current density ($J_c$) [1-12]. Various models for describing the mechanism of the formation of CDs, such as thermal spike model [13], have been discussed. The possibility of further enhancement of $J_c$ by dispersing the direction of CDs has been proposed [14]. A study on YBa$_2$Cu$_3$O$_7$ with CDs introduced in various conditions reported a correlation between the improvement of pinning properties and the amplitude of the natural splay related to the terminal dispersion of the paths of the heavy ions in the material [15]. Following this study, similar enhancements of $J_c$ by introducing splayed CDs have been reported in YBa$_2$Cu$_3$O$_7$ single crystals [16, 17] and thin films [18], and other cuprates such as Bi$_2$Sr$_2$CaCu$_2$O$_8$ single crystals [19]. Furthermore, magneto-optical imaging of a DyBa$_2$Cu$_3$O$_7$ single crystal with splayed CDs directly demonstrated the anisotropic enhancement of $J_c$ [20]. Such an enhancement of $J_c$ by introduction of CDs has also been confirmed in iron-based superconductors (IBSs) [3-12], and IBSs offer new playgrounds for vortex physics in superconductors [22-26]. Transmission electron microscope (TEM) observations showed that introduced CDs in IBSs tend to be discontinuous compared with those in cuprates [3,7,27]. The optimal splay angle for the maximum enhancement of $J_c$, determined by the competition between the beneficial effect of suppressing vortex motion by the variable inter-defect distance and forcing vortices to entangle, and the adverse effect of vortex-field misalignment, were found to be ~±5° in both YBa$_2$Cu$_3$O$_{7-\delta}$ and Ba$_{1-x}$K$_x$Fe$_2$As$_2$ [16, 21]. Moreover, in such systems with splayed CDs, an anomalous peak effect (APE) in $J_c$ at a certain magnetic field determined by the irradiation dose was observed [21, 27-31]. Since such an APE is absent in IBSs with CDs parallel to the *c*-axis and only present in those with splayed CDs when the magnetic field is applied along the *average* direction of splayed CDs, it has been proposed that it is caused by the suppression of kink motion and/or the effect of vortex entanglement in systems with splayed CDs [28]. However, it is not clear which direction is crucial to the APE, the direction of *c*-axis or the average direction of splayed CDs, since splayed CDs were always introduced *symmetrically* with respect to the *c*-axis, and therefore the average direction of splayed CDs coincided with the *c*-axis.

In the present study, we introduced splayed CDs *asymmetrically* with respect to the *c*-axis to optimally K-doped Ba-122-type IBS Ba$_{0.6}$K$_{0.4}$Fe$_2$As$_2$ single crystals, which is one of the most promising materials for the applications of superconducting tapes [32] and wires [33], by irradiating 2.6 GeV U ions. We made systematic measurements of the magnetic field dependence of $J_c$ including its angular dependence to confirm the significance of the average direction of splayed CDs and the relationship between the APE and the effect of vortex entanglement by splayed CDs.

## 2. Experimental details

$Ba_{0.6}K_{0.4}Fe_2As_2$ single crystals were synthesized by FeAs self-flux method [24, 34, 35]. Ba chunks (99.9 %), K ingots (99.5%), and FeAs powder were used as starting materials. FeAs was prepared by sealing stoichiometric amounts of As grains (7N) and Fe powder (99.9%) in an evacuated quartz tube and reacting them at 700 °C for 40 h after heating at 500 °C for 10 h. A mixture with a ratio of Ba : K : FeAs = 0.6 : 0.44 : 4 was placed in an alumina crucible in an argon-filled glove box. The alumina crucible was sealed in a stainless steel tube with a stainless steel cap [34]. The whole assembly was heated for 10 h at 1100 °C after a preliminary heating at 600 °C for 5 h, followed by cooling to 800 °C at a rate of 5 °C /h for the crystal growth. The critical temperature ($T_c$) of pristine crystals was 38.5 K±0.4 K, and the width of the transition ($\Delta T_c$) was less than 1.5 K [24]. The typical $J_c$ value of the pristine sample is 2-3 $MA/cm^2$ at 2 K under self-field [24]. The crystals were shaped into rectangular parallelepipeds with lengths and widths ~500 μm and thicknesses less than 20 μm, which is small enough compared with the projected range of 2.6 GeV U ions (65 μm).

2.6 GeV U ions were irradiated at RIKEN Ring Cyclotron in RI Beam Factory operated by RIKEN Nishina Center and CNS, The University of Tokyo. The incident directions of U ions were changed by tilting the crystals about their *ab*-plane, and the angle of CDs ($\theta_{CD}$) is denoted by the angle between their *c*-axis and the ion beam. The irradiation dose is evaluated by the dose-equivalent magnetic field called "matching field", at which all defects are occupied by single vortices;

$$B_\Phi = n\Phi_0. \qquad (1)$$

Here, $n$ is the areal density of CDs and $\Phi_0$ is a flux quantum. Schematic figures how we introduced splayed CDs asymmetrically with respect to the *c*-axis are shown in Fig. 1; (a) tilt the average direction of CDs from the *c*-axis, (b) change the irradiation dose of one set of CDs from that of another, and (c) tilt the splay plane, on which two sets of splayed CDs are residing, from the *c*-axis. Even after irradiation with total $B_\Phi = 8$ T, the superconducting transition remains sharp, and $T_c$ of the crystal was decreased only slightly to ~37.8 K, possibly due to point defects created by secondary electrons [10] as shown in Fig. 1(d). It should be noted that at a large fluence of 2.6 GeV U of $B_\Phi = 8$ T used exclusively in the present study, $J_c$ is enhanced by more than a factor of 5 compared with that in the pristine crystal. So, $J_c$ characteristics of all the irradiated crystals are governed by the pinning induced by introduced CDs, making minute variations of physical properties in the unirradiated crystals irrelevant. Furthermore, compared with the case of $Ba(Fe,Co)_2As_2$ irradiated by 200 MeV Au, where the density of CDs is ~40% of the dose of injected ions and the introduced CDs are discontinuous [3], we have confirmed that the introduced CDs are continuous and their density is nearly the same as the dose of injected 2.6 GeV U ions by the STEM observation [27]. It has also been reported that the diameter of CDs generated by 2.6 GeV U ions is 7~8 nm [27]. In addition, the variation of the diameter of CDs through the samples is negligible, since the sample is much thinner than the projection range of 2.6 GeV U ions.

Magnetization of the crystal was measured by a superconducting quantum interference device (SQUID) magnetometer (MPMS-5XL, Quantum Design). Here we define $\theta_H$ as the angle between the *c*-axis of the crystal and the external magnetic field. Magnetization measurements were performed for fields parallel to the *c*-axis ($\theta_H = 0°$) and fields tilted from the *c*-axis ($\theta_H \neq 0°$) in the splay plane. For the former measurement, the crystal was placed in a quartz sample holder and fixed with Apiezon N grease, while for the latter measurement, an acrylic sample holder with a rotating sample slot was employed. For several directions of magnetic field, the component of magnetization parallel to the field was measured. *Average* in-plane $J_c$ was calculated from the results of the magnetization measurements using the extended Bean model [36, 37]. It should be noted that in superconductors with splayed CDs, there appears in-plane $J_c$ anisotropy. At the present stage, we cannot determine which component, or both, is responsible for the anomalous peak effect. Hence, to circumvent such complexities, we strictly limit our discussion on the average in-plane $J_c$. In addition, in the case of fields tilted from the *c*-axis ($\theta_H \neq 0°$) we corrected the magnetic field by using Blatter's scaling [38];

$$\widetilde{H} = \varepsilon_\theta H \tag{2}$$

$$\varepsilon_\theta^2 = \varepsilon^2 \sin^2\theta_H + \cos^2\theta_H. \tag{3}$$

The anisotropy parameter $\varepsilon = 1/\gamma \sim 1/1.4$, typical value for IBSs at 20 K [39, 40], was used. It should be noted that the scaling of the magnetic field does not affect the results significantly as long as the anisotropy parameter is small, typically less than 2. We also corrected $J_c$ calculated from the measured magnetization, which is the component parallel to the external field, to the real in-plane $J_c$;

$$J_{c,ab} = J_c/\cos\theta_H. \tag{4}$$

This correction is based on the assumption that the supercurrent flows in the *ab*-plane and the direction of the magnetization is parallel to the *c*-axis in our thin samples and $\theta_H$ less than 45°. We confirmed the direction of the magnetization in the case of field *H* tilted from the *c*-axis by measuring two components of magnetization for one of the samples with splayed CDs at $\theta_H = 19.8°$. The configuration of the measurements is shown in Fig. 2(a), where in addition to the ordinary longitudinal component of magnetic moment $m_l$ parallel to *H*, another transverse component $m_t$ perpendicular to *H* was measured. The hysteresis loops for $m_l$ and $m_t$ are shown in Fig. 2(b) and (c), respectively. From the measured $m_l$ and $m_t$, the angle of magnetization from the direction of *H*, $\tan^{-1}(m_t/m_l)$ can be calculated. The calculated angle as a function of *H* is shown in Fig. 2(d), which indicates that the average direction of magnetization is 19.8±1.6° from the direction of *H*, or equivalently 0.0±1.6° from the *c*-axis, which justifies our analyses. It should be noted that these corrections to the magnetic field in the case of field tilted from the *c*-axis do not affect the comparison of the magnitude of the APE for various field directions discussed below.

### 3. Experimental Results

The first case of introducing CDs asymmetrically with respect to the *c*-axis is tilting the average direction of CDs from the *c*-axis, as shown in Fig. 1(a). Figure 3 shows the magnetic field dependence of $J_c$ at various temperatures of Ba$_{0.6}$K$_{0.4}$Fe$_2$As$_2$ that are irradiated by two sets of 2.6 GeV U ions each with $B_\Phi$ = 4 T at (a) $\theta_{CD}$ = ±15° (symmetric with respect to the *c*-axis), (b) $\theta_{CD}$ = 10°±15° ( = +25° and -5°), and (c) $\theta_{CD}$ = 20°±15° ( = +35° and +5°) for fields parallel to the *c*-axis ($\theta_H$ = 0°). It should be noted that data in the range of the return branch is omitted to show the intrinsic magnetic field dependence of $J_c$. The APE is observed as a local maximum of $J_c$ at around 30 kOe, close to 1/3$B_\Phi$. At low fields, changes in the slope of the magnetic-field dependence of $J_c$ due to the self-field effect can be observed as shown by red arrows in Fig. 3 (c) [6, 41, 42], which confirms that the pinning mechanism in our samples is dominated by correlated pinning centers. The APE shows up more clearly at higher temperatures. Figure 3 (d) shows magnetic field dependence of $J_c$ at 20 K of the above three cases. As the average directions of CDs are tilted from the *c*-axis, the APE is gradually suppressed. This result supports the relationship between the APE and the effect of vortex entanglement by splayed CDs. Namely, when the field is applied parallel to the *c*-axis to samples with splayed CDs whose average direction is tilted from the *c*-axis, one set of CDs at $\theta_{CD}$ closer to 0° (the direction of the *c*-axis and applied field) is preferentially occupied and thus the effect of vortex entanglement is suppressed. On the other hand, if the suppression of kink motion is the main mechanism for the APE, the average direction of the CDs would not play an important role, since the kink motion should be suppressed in a wide range of field direction in the presence of splayed CDs.

Figure 4 (a) shows the magnetic field dependence of $J_c$ at 20 K of Ba$_{0.6}$K$_{0.4}$Fe$_2$As$_2$ that is irradiated by 2.6 GeV U ions with $B_\Phi$ = 4 T + 4 T and $\theta_{CD}$ = 20°±15° ( = +35° and +5°) for fields parallel to the *c*-axis ($\theta_H$ = 0°) and fields tilted from the *c*-axis ($\theta_H \neq 0°$) in the splay plane, on which two sets of splayed CDs are residing, as shown in Fig. 4 (c). Figure 4 (b) shows the $\theta_H$ dependence of $J_c$ at 20 K and 27.5 kOe extracted from Fig. 4 (a). Both in Figs. 4(a) and (b), we corrected the magnetic field to $\tilde{H}$ and $J_c$ to $J_{c,ab}$ in the case of fields tilted from the *c*-axis ($\theta_H \neq 0°$). The blue and red broken lines show angles for two sets of columnar defects ($\theta_{CD}$ and $\theta_{CD}$') and their average ($\theta_{CD}^{av}$), respectively. As the applied field is tilted towards the average direction of splayed CDs, $\theta_{CD}^{av}$, the suppressed APE is restored gradually, which confirms the significance of the average direction of splayed CDs for the APE.

The second case of introducing CDs asymmetric with respect to the *c*-axis is making the irradiation dose of one set of CDs different from that of another, as shown in Fig. 1(b). Figure 5 shows the magnetic field dependence of $J_c$ at various temperatures of $Ba_{0.6}K_{0.4}Fe_2As_2$ that are irradiated by 2.6 GeV U ions with $\theta_{CD} = \pm 15°$ and (a) $B_\Phi = 4$ T + 4 T (symmetric with respect to the *c*-axis), (b) $B_\Phi = 3.5$ T + 4.5 T, and (c) $B_\Phi = 2$ T + 6 T. The APE at around 30 kOe, close to 1/3 $B_\Phi$, is observed most clearly in the symmetric case, especially at high temperatures. At low temperatures, however, it is not so clear due to large irreversible magnetization. How the asymmetric doses for two sets of CDs affect the APE is shown in Fig. 5(d) by comparing the $J_c$-$H$ data at 20 K. As the irradiation doses of two sets of CDs are unbalanced, the APE is suppressed. This result again supports the scenario that the APE is caused by the entanglement of vortices by splayed CDs, since the imbalance of the irradiation dose of each CD reduces the number of intersections of CDs, and thus the effect of vortex entanglement is suppressed.

The third case of introducing CDs asymmetric with respect to the *c*-axis is tilting the splay plane from the *c*-axis. The angle of the splay plane ($\theta_{SP}$) is defined by the angle between the *c*-axis and the splay plane spanned by two sets of CDs, as shown in Fig. 1(c). Figure 6 shows the magnetic field dependence of $J_c$ at various temperatures of $Ba_{0.6}K_{0.4}Fe_2As_2$ that are irradiated by 2.6 GeV U ions with $B_\Phi = 4$ T + 4 T, $\theta_{CD} = \pm 15°$, and (a) $\theta_{SP} = 0°$ (symmetric with respect to the *c*-axis), (b) $\theta_{SP} = 10°$, (c) $\theta_{SP} = 20°$, and (d) $\theta_{SP} = 30°$. Magnetic field dependence of $J_c$ at 5 K and 20 K of the above four cases are summarized in Figs. 6(e) and (f), respectively. At 5 K, as $\theta_{SP}$ becomes large, the magnetic field dependence of $J_c$ becomes stronger without any peak at intermediate fields. At 20 K, as the splay plane is tilted from the *c*-axis, the APE is strongly suppressed. These results once again support the relationship between the APE and the effect of vortex entanglement by splayed CDs. When the field is applied parallel to the *c*-axis to the sample with splayed CDs with their splay plane tilted from the *c*-axis, the misalignment between the field and CDs weakens the pinning effect of CDs, and thus the effect of vortex entanglement is suppressed.

Unlike the systematic $\theta_{SP}$ dependence of $J_c$ at high fields, $\theta_{SP}$ dependence of $J_c$ close to zero field is complicated as shown in Figs. 6(e) and (f). We believe that at a fixed density of CDs as in our measurements, $J_c$ at zero field should not change too much depending on the configuration of CDs as long as their directions are not too much different. So, the apparent complicated $\theta_{SP}$ dependence of $J_c$ close to zero field is an artifact due to the presence of macroscopic defects in the crystal and uncertainty in the estimation of the sample thickness, both of which affect the evaluation of $J_c$. It should be noted, however, that such uncertainty in the absolute value of $J_c$ does not affect the behavior of APE as a function of $\theta_{SP}$.

Figure 7(a) shows the magnetic field dependence of $J_c$ at 20 K of $Ba_{0.6}K_{0.4}Fe_2As_2$ that is irradiated by 2.6 GeV U ions with $B_\Phi$ = 4 T + 4 T, $\theta_{CD} = \pm 15°$ and $\theta_{SP} = 20°$ for fields parallel to the c-axis ($\theta_H = 0°$) and fields tilted from the c-axis ($\theta_H \neq 0°$) in a plane perpendicular to the splay plane, as shown in Fig. 7(c). The $\theta_H$ dependence of $J_c$ at 20 K and 27.5 kOe extracted from Fig. 7(a) is plotted in Fig. 7(b). Similar to the case of Fig. 4, we corrected the magnetic field to $\widetilde{H}$ and $J_c$ to $J_{c,ab}$ in the case of fields tilted from the c-axis ($\theta_H \neq 0°$). As the applied field is tilted towards the splay plane, the APE is restored, which also confirms the significance of the average direction of splayed CDs for the APE.

## 4. Discussion

When we discuss the APE in systems with splayed CDs, two important values are the peak value of $J_c$, which indicates the peak strength, and the value of magnetic field corresponding to the peak. In terms of the peak strength, our results in the first and third experiments mentioned above show that the APE appears strongly when the direction of the applied field is close to the average direction of CDs, regardless of the condition whether CDs are introduced symmetrically with respect to the c-axis or not. When the two sets of CDs have the same pinning strengths, the effect of vortex entanglement should be the strongest when the field is applied along the average direction of CDs. Therefore, the results of the first and third experiments support our assertion that the APE is caused by the effect of vortex entanglement. In the case of $YBa_2Cu_3O_7$ thin films with splayed CDs symmetric with respect to the c-axis, $J_c$ shows a peak when the magnetic field is applied along the c-axis, which is the average direction of CDs, as long as $|\theta_{CD}|$ is not very large [43].

However, as shown in Fig. 4(b), the value of $\theta_H$ where the strongest APE appears shifts slightly from the average direction of splayed CDs towards the c-axis. In systems with only one set of CDs tilted from the c-axis, asymmetric angular dependences with respect to the direction of CDs were reported in Josephson plasma-resonance experiments of $Bi_2Sr_2CaCu_2O_{8+y}$ single crystals [44]. In addition, in the low-field region, where $4\pi|M|$ is comparable to the external field $|H|$, the peak shifts from the direction of CDs towards either the c-axis or the ab-planes in the angular dependence of the irreversible magnetization were reported in $YBa_2Cu_3O_7$ and $2H-NbSe_2$ single crystals with CDs tilted away from the c-axis [45]. Such shifts can occur due to the misalignment between the external and internal field directions caused by the competition between anisotropy and geometry effects. However, in the high-field region close to 30 kOe as in our case, it is unreasonable for the internal field direction to be tilted away from the external field direction. It should be noted, however, that in systems with two sets of CDs whose average direction is tilted from the c-axis as in our case, the angular dependence of physical properties has not been clearly understood.

The fact that the peak of $J_c$ shifts from $\theta_{CD}^{av}$ towards the $c$-axis can be explained by combining two considerations. First, since the APE occurs only in superconductors with splayed CDs, occurrence of vortex entanglement is essential. Vortex entanglement should occur most frequently when two sets of CDs are equally occupied by vortices. Second, the pinning force of vortices in layered superconductors with CDs per superconducting layer is given by $U_p/\xi_{ab}$ ($U_p$: pinning potential, $\xi_{ab}$: in-plane coherence length), where $U_p$ is proportional to the cross section of CDs in a layer. Since the diameter of CDs perpendicular to their length does not depend on the direction of CDs, $\theta_{CD}$, the cross section of CDs in a layer is proportional to $1/\cos\theta_{CD}$, and hence the pinning force increases with $\theta_{CD}$. The situations for different $\theta_H$ are schematically shown in Figs. 4(d) and (e). Here, two sets of CDs with $\theta_{CD}$ and $\theta_{CD}$' are shown by tilted thick orange lines and the depth of orange color is proportional to the trapping probability of vortices on CDs. A representative superconducting layer is also shown by a thick blue horizontal line. When the magnetic field is applied at $\theta_H = \theta_{CD}^{av}$ as shown in Fig. 4(d), due to the larger pinning force of CDs at $\theta_{CD}$' with larger angle, they trap more vortices, making the probability of vortex entanglement less. On the other hand, when the magnetic field is applied at $\theta_H = \theta_H^{peak}$ ($<\theta_{CD}^{av}$), the vortex entanglement should occur most frequently since two sets of CDs trap vortices with equal probabilities as shown in Fig. 4(e).

In terms of the value of magnetic field corresponding to the peak, the APE appears at around 30 kOe, roughly $1/3$ $B_\Phi$, which can be determined by the competition between the vortex-defect and the vortex-vortex interactions [46, 47]. It should be noted that as in the case of Josephson plasma resonance [3031, 43] and $c$-axis transport [48] in $Bi_2Sr_2CaCu_2O_{8+y}$, the exact value of this fraction can be unimportant, and it may vary from $\sim 1/5$ to $\sim 1/3$. In the second experiment, by changing the irradiation dose of one set of CDs from that of another, the reduction of the number of intersections of CDs may change the condition of the competition, which can not only suppress the APE but also change the value of magnetic field corresponding to the APE. As the asymmetry in doses is increased, we observe a weak shift of the peak field to lower fields as well as the suppression of the APE. From Fig. 5(d), the value of magnetic field corresponding to the APE seems to shift from 26 kOe for $B_\Phi = 4$ T + 4 T to 21 kOe for $B_\Phi = 2$ T + 6 T, as the irradiation doses of two sets of CDs are unbalanced. Such a shift can be explained by considering the "asymmetric doses" as "symmetric doses" + "additional dose". In such a scenario, we may expect that the APE appears close to $1/3$ of the symmetric part of $B_\Phi$, which is smaller than $1/3$ of total $B_\Phi$. However, this shift is not so clear and in this study we have limited data; the total irradiation doses of two sets of CDs are always 8 T. Thus, in order to confirm the relationship between the value of magnetic field corresponding to the peak and the effect of vortex entanglement, which depends on the number of intersections of CDs, further studies with much different irradiation conditions, such as $B_\Phi$ = 0.5 T + 7.5 T or different total matching fields, are needed.

5. **Summary**

Magnetic field dependence of $J_c$ is systematically studied in $Ba_{0.6}K_{0.4}Fe_2As_2$ single crystals with asymmetric splayed CDs with respect to the *c*-axis introduced by irradiating 2.6 GeV U ions. By tilting the average direction of CDs from the *c*-axis, changing the irradiation dose of one set of CDs from that of another, and tilting the splay plane from the *c*-axis, the anomalous peak effect (APE) that appears around 1/3 $B_\Phi$ is suppressed. These results indicate that the APE is caused by the effect of vortex entanglement rather than the suppression of kink motion in superconductors with splayed CDs. In the first and third cases, the APE is restored by tilting the applied field towards the average direction of splayed CDs. These results confirm the significance of the average direction of splayed CDs for the APE. The slight deviation of optimum $\theta_H$ for the strongest APE from the average direction of splayed CDs can be explained by considering the enhanced pinning effects of CDs at larger $\theta_{CD}$.

**Acknowledgement**

This work was partially supported by a Grant-in-Aid for Scientific Research (A) (17H01141) by the Japan Society for the Promotion of Science (JSPS). We thank T. Nojima for technical assistance.

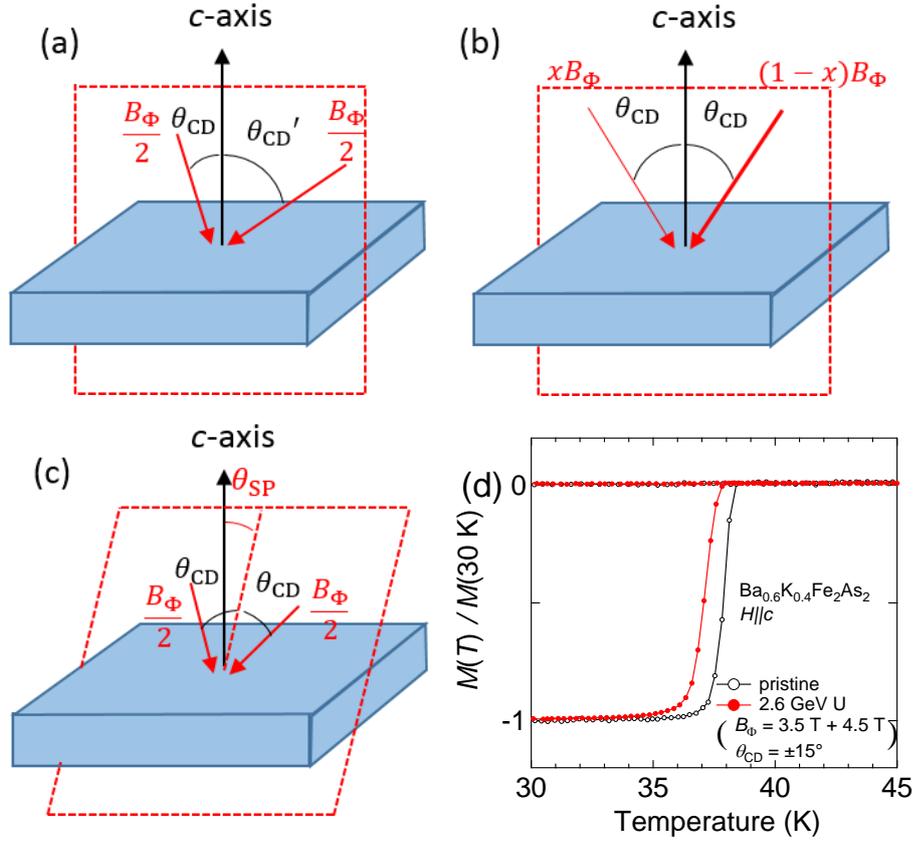

Fig. 1. (a)-(c) Three configurations of splayed CDs introduced asymmetrically with respect to the *c*-axis by irradiating 2.6 GeV U ions. (a) The case of tilting the average direction of CDs from the *c*-axis, (b) the case of changing the irradiation dose of one set of CDs from that of another, and (c) the case of tilting the splay plane from the *c*-axis. (d) Temperature dependence of magnetization for a pristine $Ba_{0.6}K_{0.4}Fe_2As_2$ ($H$ = 10 Oe) and a irradiated $Ba_{0.6}K_{0.4}Fe_2As_2$ ($H$ = 5 Oe) that is irradiated by 2.6 GeV U ions with total $B_\Phi$ = 8 T ($B_\Phi$ = 3.5 T + 4.5 T, $\theta_{CD}$ = ±15°).

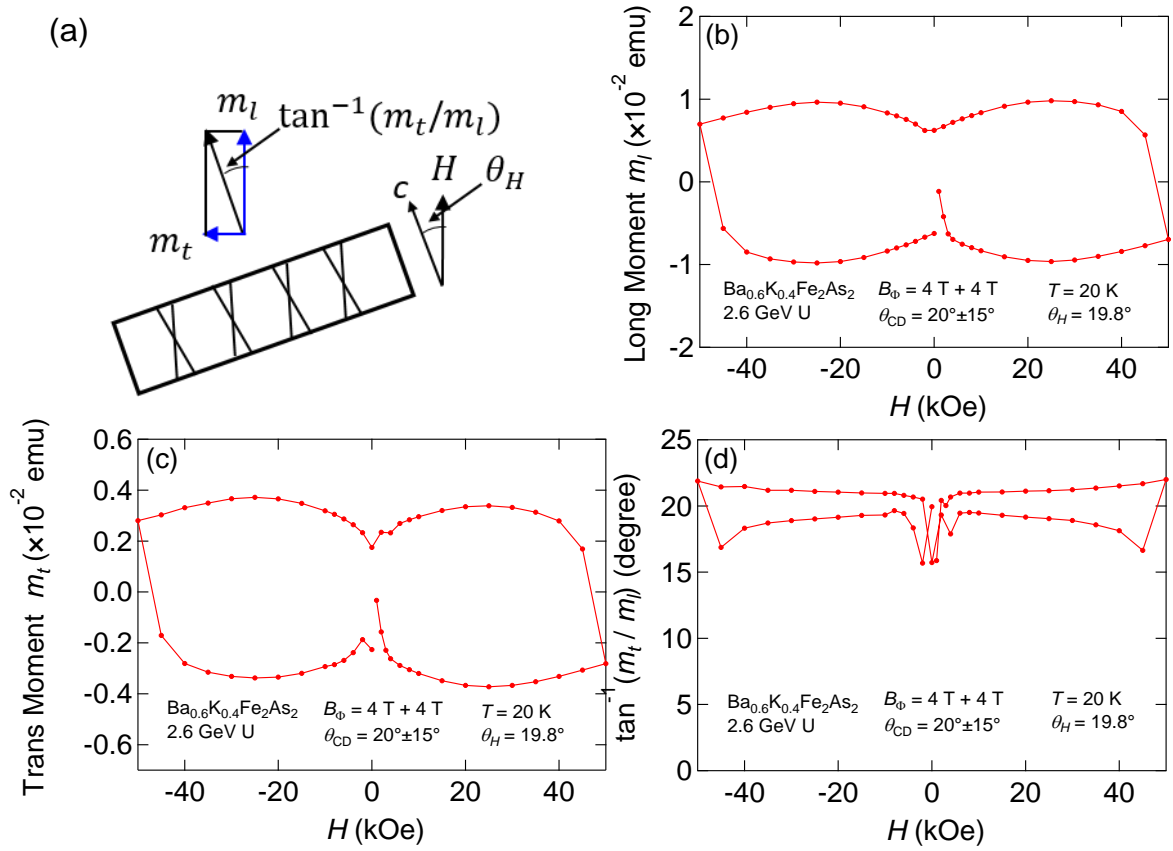

Fig. 2. (a) The configuration of the sample with splayed CDs and external magnetic field, $H$. $\boldsymbol{m}_l$ and $\boldsymbol{m}_t$ indicate longitudinal and transverse magnetization vector of the sample. Hysteresis loops of (b) longitudinal and (c) transverse components of magnetization and (d) the direction of magnetization calculated from the ratio of the two components at 20 K for $Ba_{0.6}K_{0.4}Fe_2As_2$ that is irradiated by 2.6 GeV U ions with $B_\Phi = 4$ T + 4 T and $\theta_{CD} = 20°\pm15°$ (= +35° and +5°) at $\theta_H = 19.8°$. The fact that the angle of magnetization with respect to the external field, $\tan^{-1}(m_t/m_l)$, at $\theta_H = 19.8°$ is ~20° indicates that the direction of the magnetization is almost parallel to the $c$-axis.

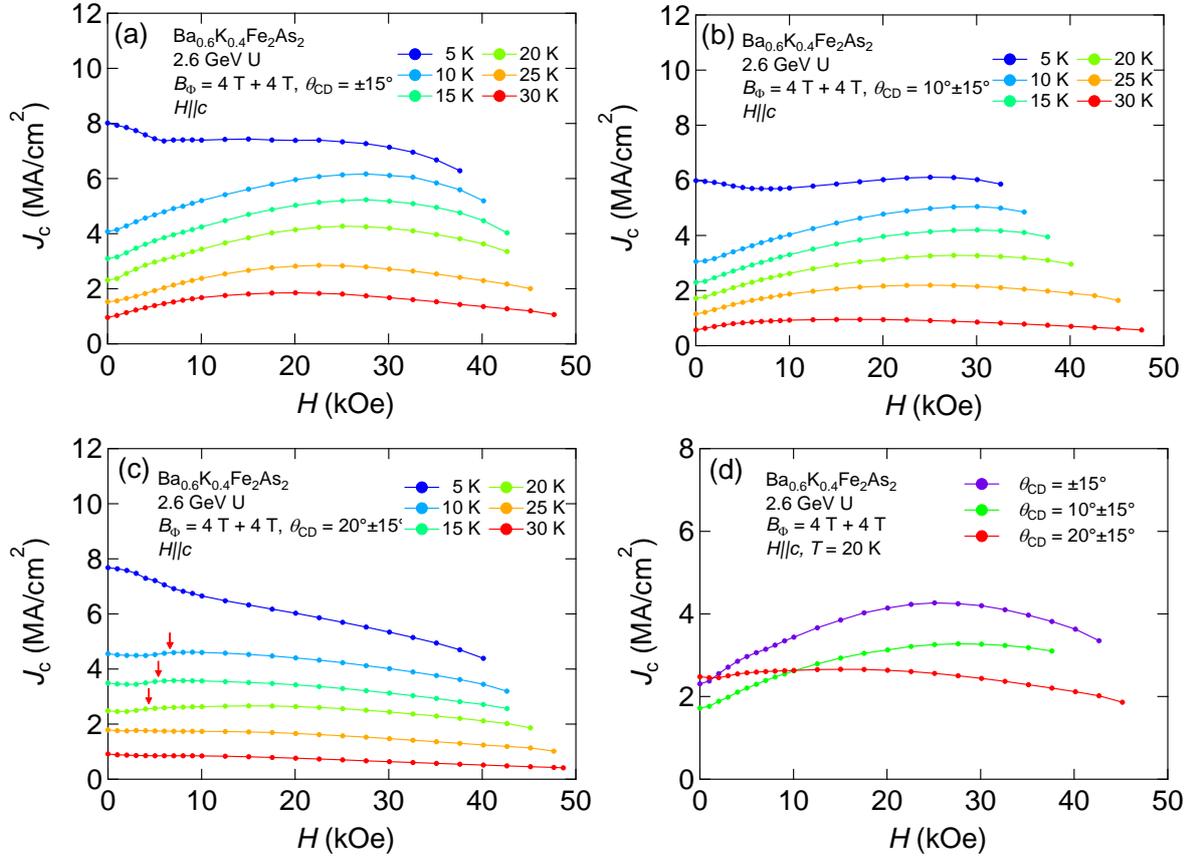

Fig. 3. Magnetic field dependence of $J_c$ at various temperatures of $Ba_{0.6}K_{0.4}Fe_2As_2$ that are irradiated by 2.6 GeV U ions with $B_\Phi = 4\ \text{T} + 4\ \text{T}$ and (a) $\theta_{CD} = \pm15°$ (symmetric with respect to the $c$-axis), (b) $\theta_{CD} = 10°\pm15°$ ( $= +25°$ and $-5°$), and (c) $\theta_{CD} = 20°\pm15°$ ( $= +35°$ and $+5°$) for fields parallel to the $c$-axis ($\theta_H = 0°$). (d) Magnetic field dependence of $J_c$ at 20 K of the above three cases. Red arrows in (c) show examples of the onset of self-field effect related to the suppression of $J_c$ due to the curvature of vortices near zero field [6].

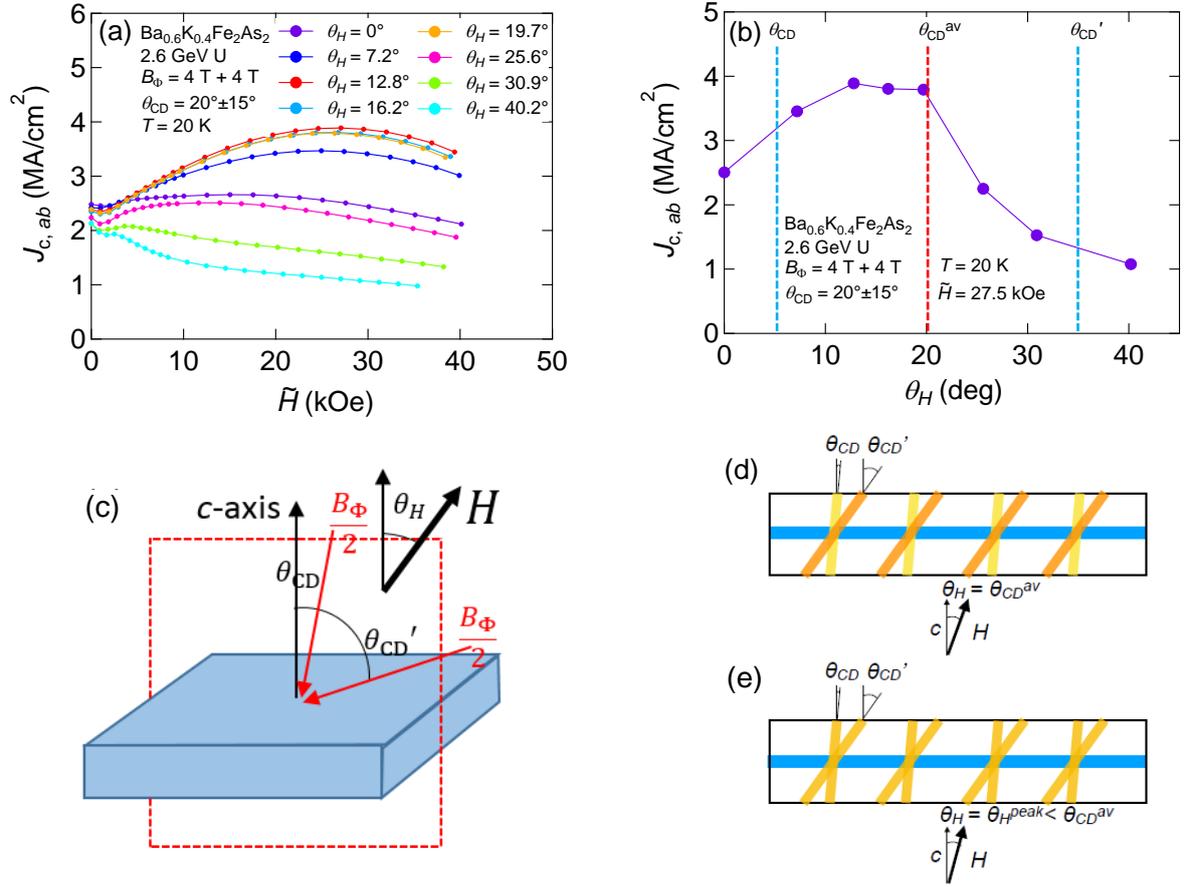

Fig. 4. (a) In-plane critical current density $J_{c,ab}$ as a function of applied field rescaled by Blatter's anisotropic scaling $\tilde{H}$ at 20 K of $Ba_{0.6}K_{0.4}Fe_2As_2$ that is irradiated by 2.6 GeV U ions with $B_\Phi = 4$ T + 4 T and $\theta_{CD} = 20° \pm 15°$ (= +35° and +5°) for various $\theta_H$. (b) $\theta_H$ dependence of $J_{c,ab}$ at 20 K and $\tilde{H} = 27.5$ kOe taken from (a). The blue and red broken lines in (b) show angles for two sets of columnar defects ($\theta_{CD}$ and $\theta_{CD}'$) and their average ($\theta_{CD}^{av}$), respectively. (c) The configuration of splayed CDs at $\theta_{CD}$ and $\theta_{CD}'$ with the direction of the magnetic field. Trapping probabilities of vortices by two sets of CDs at $\theta_{CD}$ and $\theta_{CD}'$ when $H$ is applied at (d) $\theta_H = \theta_{CD}^{av}$ and (e) $\theta_H = \theta_H^{peak}$ ($< \theta_{CD}^{av}$) are schematically shown. The depth of orange color is proportional to the trapping probability of vortices. The thick blue horizontal line schematically shows one superconducting layer to highlight the difference in the cross section of CDs across it.

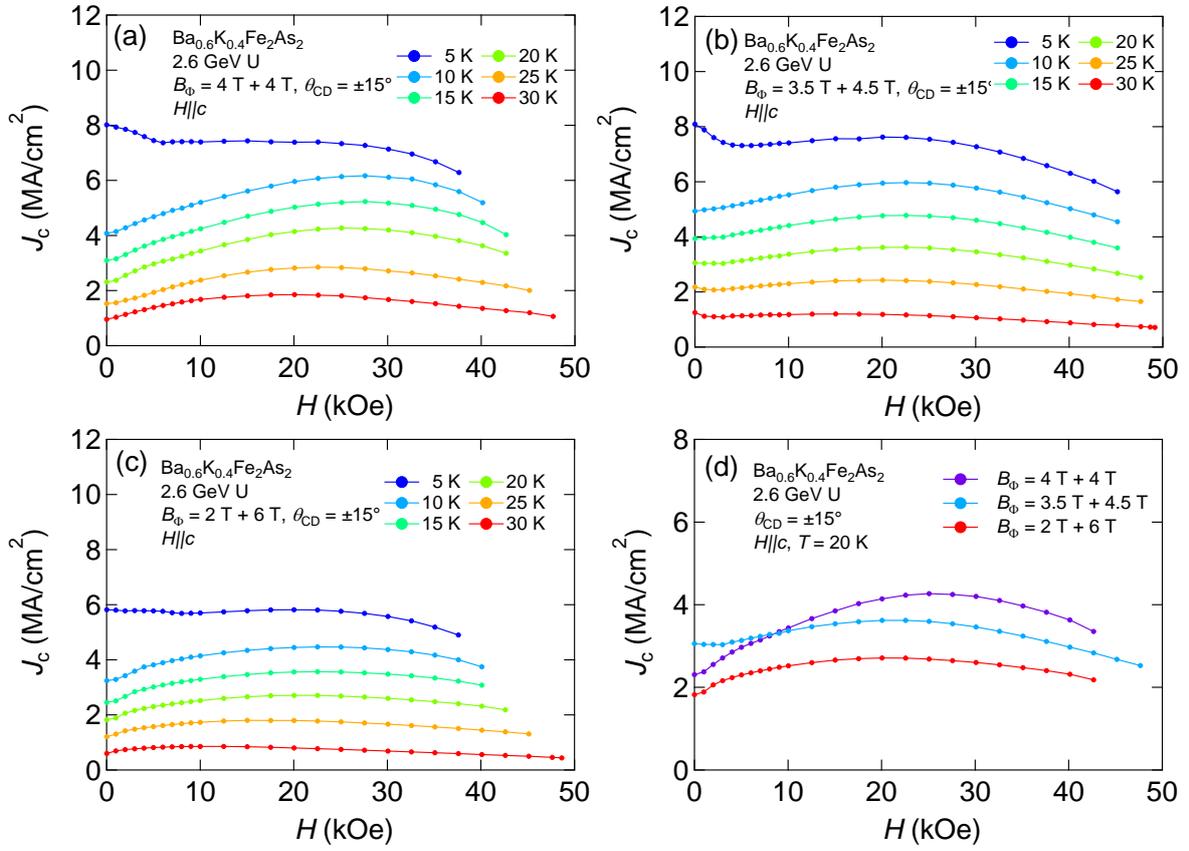

Fig. 5. Magnetic field dependence of $J_c$ at various temperatures of $Ba_{0.6}K_{0.4}Fe_2As_2$ that are irradiated by 2.6 GeV U ions at $\theta_{CD} = \pm 15°$ and (a) $B_\Phi = 4\,T + 4\,T$ (symmetric with respect to the $c$-axis), (b) $B_\Phi = 3.5\,T + 4.5\,T$, and (c) $B_\Phi = 2\,T + 6\,T$. (d) Magnetic field dependence of $J_c$ at 20 K of the above three cases.

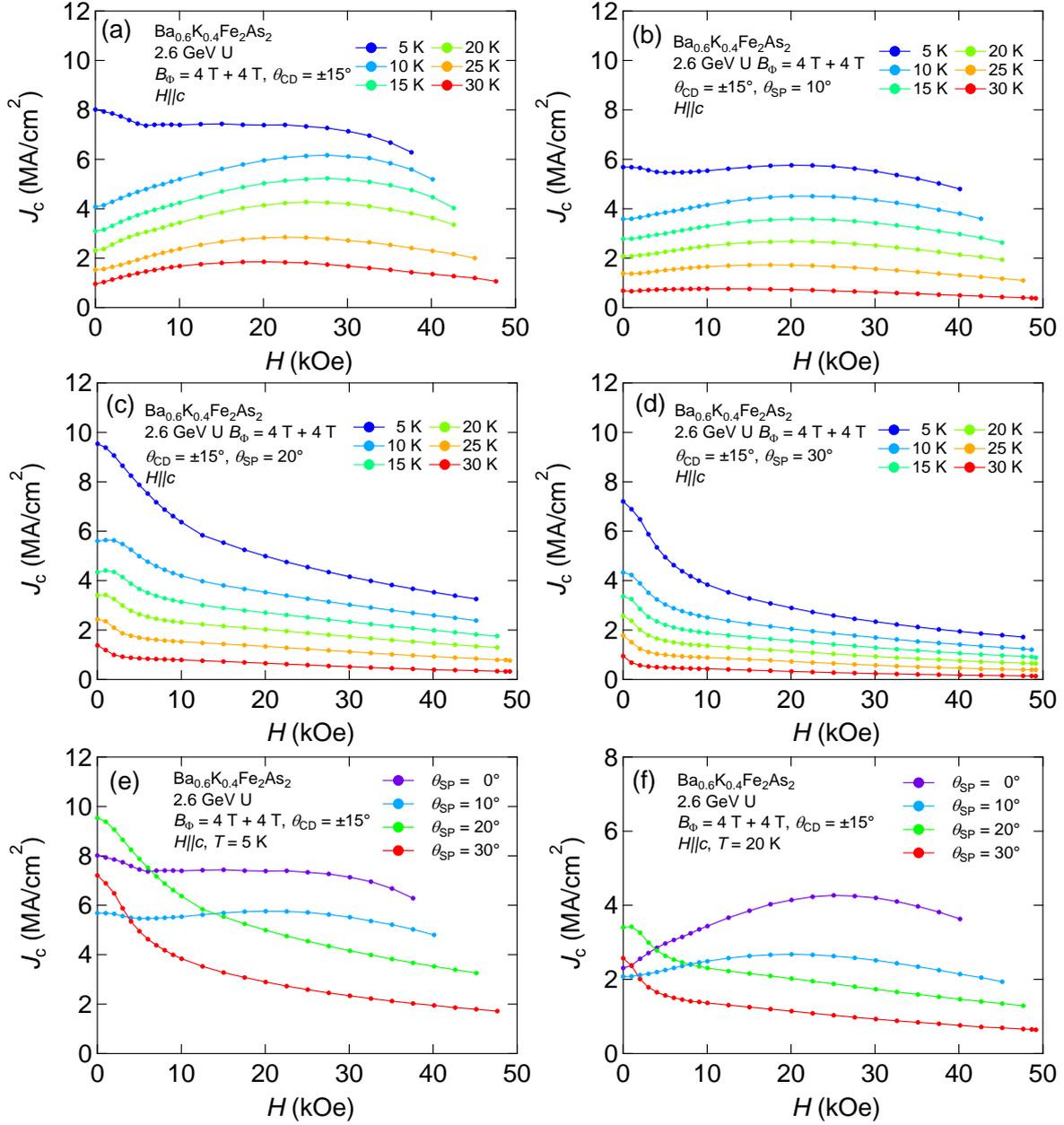

Fig. 6. Magnetic field dependence of $J_c$ at various temperature of $Ba_{0.6}K_{0.4}Fe_2As_2$ that are irradiated by 2.6 GeV U ions with $B_\Phi$ = 4 T + 4 T, $\theta_{CD}$ = ±15° and (a) $\theta_{SP}$ = 0° (symmetric with respect to the $c$-axis), (b) $\theta_{SP}$ = 10°, (c) $\theta_{SP}$ = 20°, and (d) $\theta_{SP}$ = 30°. Magnetic field dependence of $J_c$ at (e) 5 K and (f) 20 K of the above four cases.

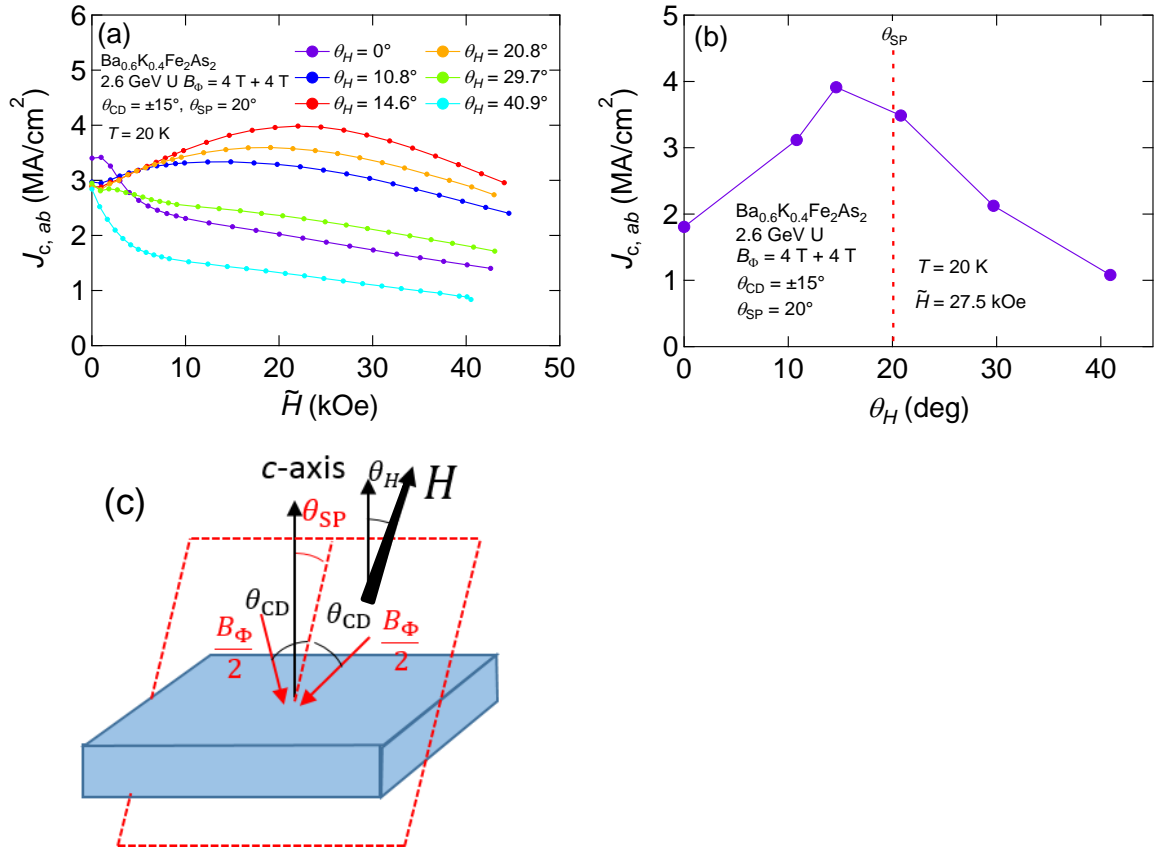

Fig. 7. (a) In-plane critical current density $J_{c,ab}$ as a function of applied field rescaled by Blatter's anisotropic scaling $\tilde{H}$ at 20 K of $Ba_{0.6}K_{0.4}Fe_2As_2$ that is irradiated by 2.6 GeV U ions with $B_\Phi = 4$ T + 4 T, $\theta_{CD} = \pm 15°$ and $\theta_{SP} = 20°$ for various $\theta_H$. (b) $\theta_H$ dependence of $J_{c,ab}$ at 20 K and $\tilde{H} = 27.5$ kOe taken from (a). (c) The configuration of splayed CDs with the direction of the magnetic field, which is tilted from the c-axis towards the splay plane.